\documentclass[a4paper,UKenglish,cleveref, autoref, thm-restate]{lipics-v2021}

\usepackage{amssymb}
\usepackage{amsmath}
\usepackage{pifont}
\newcommand{\cmark}{\ding{51}}
\newcommand{\xmark}{\ding{55}}

\usepackage{todonotes}
\usepackage{booktabs}
\usepackage{multirow}
\usepackage[algo2e,vlined]{algorithm2e}

\nolinenumbers
\hideLIPIcs

\title{A Fast and Tight Heuristic for A* in Road Networks}
\author{Ben Strasser}{Germany}{academia@ben-strasser.net}{}{}
\author{Tim Zeitz}{Karlsruhe Institute of Technology, Germany}{tim.zeitz@kit.edu}{https://orcid.org/0000-0003-4746-3582}{}

\authorrunning{B. Strasser and T. Zeitz}
\Copyright{Ben Strasser and Tim Zeitz}



\EventEditors{David Coudert and Emanuele Natale}
\EventNoEds{2}
\EventLongTitle{19th International Symposium on Experimental Algorithms (SEA 2021)}
\EventShortTitle{SEA 2021}
\EventAcronym{SEA}
\EventYear{2021}
\EventDate{June 7--9, 2021}
\EventLocation{Nice, France}
\EventLogo{}
\SeriesVolume{190}
\ArticleNo{6}

\begin{document}

\maketitle

\begin{abstract}
We study exact, efficient and practical algorithms for route planning in large road networks.
Routing applications often require integrating the current traffic situation, planning ahead with traffic predictions for the future, respecting forbidden turns, and many other features depending on the exact application.
While Dijkstra's algorithm can be used to solve these problems, it is too slow for many applications.
A* is a classical approach to accelerate Dijkstra's algorithm.
A* can support many extended scenarios without much additional implementation complexity.
However, A*'s performance depends on the availability of a good heuristic that estimates distances.
Computing tight distance estimates is a challenge on its own.
On road networks, shortest paths can also be quickly computed using hierarchical speedup techniques.
They achieve speed and exactness but sacrifice A*'s flexibility.
Extending them to certain practical applications can be hard.
In this paper, we present an algorithm to efficiently extract distance estimates for A* from Contraction Hierarchies (CH), a hierarchical technique.
We call our heuristic CH-Potentials.
Our approach allows decoupling the supported extensions from the hierarchical speed-up technique.
Additionally, we describe A* optimizations to accelerate the processing of low degree nodes, which often occur in road networks.
\end{abstract}

\section{Introduction}
\label{sec:intro}
The past decade has seen a plethora of research on route planning in large street networks~\cite{bdgmpsww-rptn-16}.
Routing a user through a road network can be formalized as the shortest path problem in weighted graphs.
Nodes represent intersections.
Roads are modeled using edges.
Edges are weighted by their traversal times.
The problem can be solved with Dijkstra's algorithm~\cite{d-ntpcg-59}.
Unfortunately, on continental sized networks, it is too slow for many applications.
Thus, speed-up techniques have been developed.
One popular example are Contraction Hierarchies~(CH)~\cite{gssv-erlrn-12}.
They have been used successfully in many real world applications.
A CH exploits the inherent hierarchy of road networks.
In a preprocessing step, additional shortcut edges are inserted, which allow skipping unimportant parts of the network at query time.
Another popular example is Multi-Level-Dijkstra~(MLD)~\cite{swz-umlgt-02} also known as CRP~\cite{dgpw-crprn-13}.
It is also used in practice~\cite{bingblog}.
MLD also uses shortcut edges.
Both approaches achieve speed-ups of at least three orders of magnitude over Dijkstra's algorithm.

\begin{figure}
\centering
\includegraphics[width=.6\columnwidth]{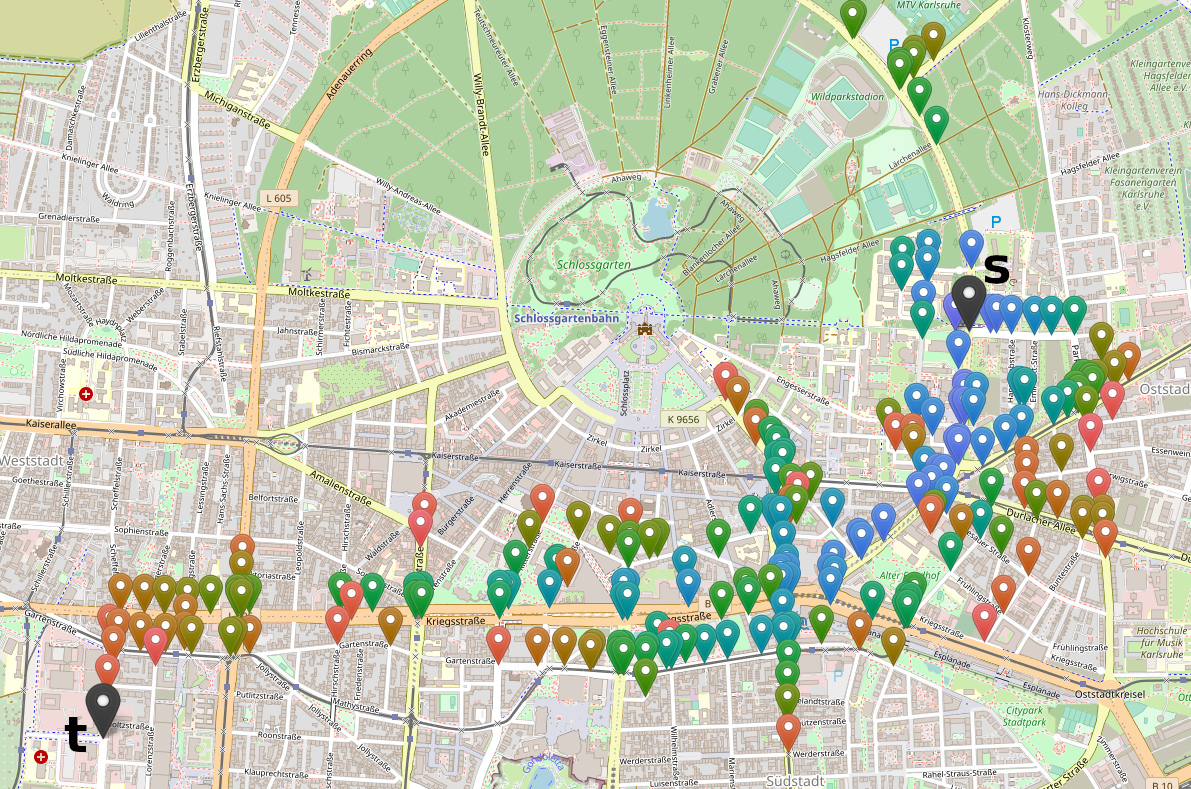}
\caption{Nodes explored by A*. Color indicates the node removal order from the queue. Blue was removed first. Next is green. Red was removed last.}
\label{img:search-space}
\end{figure}

Unfortunately, for many real world applications, this basic graph model is too simplistic.
For realistic routing, many additional features need to be considered.
This includes turn costs and restrictions, live traffic, user preferences, and traffic predictions. 
Some applications may have additional application-specific requirements.
Extending Dijkstra's algorithm to support these features is usually easy.
Extending hierarchical speed-up techniques is also possible.
However, the algorithm development is vastly more complex.
For every feature, dedicated research paper(s) exist that extend CH.
Supporting the combination of several features is even harder.
For example, we are not aware of any work combining all features mentioned above.
In this paper, we describe an algorithmic building block, that allows handling the combination of all above mentioned features -- and probably more.

Our approach decouples extensions from the hierarchical speed-up technique by utilizing the A* algorithm~\cite{hnr-afbhd-68}.
A* is a goal-directed variant of Dijkstra's algorithm.
See Figure~\ref{img:search-space} for an example of nodes traversed during an A* search.
A* uses a \emph{heuristic} to guide the search towards the goal.
A heuristic is function that maps a node $v$ onto an estimate of the distance from $v$ to the goal.
A*'s running time crucially depends on how tight this estimate is.
Further, evaluating the heuristic must be fast.
In this paper, we describe CH-Potentials, a fast heuristic with tight estimates.
Internally, the heuristic uses a CH.
Fortunately, this is an implementation detail from the perspective of the A*.
To support a new feature, we only need to modify the A* algorithm.
The heuristic core containing the CH remains untouched.
Extending A* is vastly easier than extending a CH.
This enables us to design algorithms for a multitude of features.
In addition, we describe query optimizations for handling of low-degree nodes, common in road networks.
These low degree optimizations are applicable to Dijkstra's algorithm and A*.

The rest of the paper is organized as follows.
In Section~\ref{sec:related_work}, we discuss related works on goal directed search and extensions for realistic applications for hierarchical techniques.
CH-Potentials, our new distance estimation function is introduced in Section~\ref{sec:main-algo}.
Section~\ref{sec:low-deg-improvment} discusses our improvements for the handling of low-degree nodes.
In Section~\ref{sec:extensions}, we demonstrate CH-Potential's flexibility, by describing how to apply the approach to different practical applications.
Finally, in Section~\ref{sec:experiments}, we present an experimental evaluation of our approach.

\section{Related Work}\label{sec:related_work}

There is a lot of work that extends hierarchical speed-up techniques to more complex settings~\cite{bdgmpsww-rptn-16}.
For example, in~\cite{gv-errnt-11} turn information is integrated into CH.
A considerable amount of research and engineering effort has been put into studying the combination of traffic predictions with CH.
Several papers~\cite{bdsv-tdch-09,bgns-tdcha-10,klsv-dtdch-10,bgsv-mtdtt-13} and an entire dissertation~\cite{b-tdrpc-14} have been published on the subject.
Different variants with trade-offs regarding exactness, query speed and space consumption were proposed~\cite{bgsv-mtdtt-13}.
Recently, a new approach has been published~\cite{swz-sfert-20} which simultaneously achieves competitive results in all three aspects but only at the cost of considerable implementation complexity.
CRP~(Customizable Route Planning)~\cite{dgpw-crprn-13} is an engineered variant of MLD~\cite{swz-umlgt-02} which was developed to allow updating weights without invalidating the entire preprocessing.
For this, a faster, second preprocessing phase is introduced.
It can be run regularly to update weights.
In theory, this enables the integration of live traffic and user preferences.
In practice, live traffic feed data is imperfect.
Computing ``good'' routes without undesired detours due to artifacts in the data requires additional algorithmic extensions~\cite{dss-tarrn-18}.
CRP also supports turn costs.
Integrating traffic predictions into CRP was studied in~\cite{bdpw-dtdrp-16}.
On continental sized networks, TD-CRP can only compute approximate shortest distances (rather than paths).
In~\cite{dsw-cch-15}, CH is extended to Customizable CH (CCH).
CCH also has a second preprocessing phase where weights can be altered.
Supporting turn costs in CCH was studied in~\cite{bwzz-cchtc-20}.
Other extensions studied include electric vehicle routing~\cite{DBLP:journals/algorithmica/BaumDPSWZ20,DBLP:conf/aaai/EisnerFS11} or multi-criteria optimization~\cite{fns-opca-14,gks-rpfof-10}.
While these works show that it is possible to extend hierarchical approaches, they also show that it is non-trivial.
Further, in every extension the flexibility available at query time is fairly limited.
Combining these hierarchical extensions is an unsolved problem.

CH-Potentials is not the first work to combine hierarchical approaches and A*~\cite{bdsssw-chgds-10,gkw-blwr-07,bdgwz-sfpcs-19}.
However, previous works mostly focused on accelerating hierarchical approaches further rather than exploiting A*'s flexibility.

ALT~\cite{gh-cspas-05,gw-cppsp-05} and CPD-Heuristics~\cite{DBLP:conf/ijcai/BonoGHS19} are the two techniques with high conceptual similarity to CH-Potentials.
ALT has been combined with shortcuts~\cite{bdsssw-chgds-10} and also extended for dynamic graphs~\cite{dw-lbrdg-07} and time-dependent routing~\cite{ndls-bastd-12,dn-crdtd-12}.
CPD-Heuristics are a combination of A* and Compressed Path Databases (CPD).
A CPD can quickly compute the first edge of a shortest path between any two nodes.
In~\cite{DBLP:conf/ijcai/BonoGHS19}, SRC~\cite{DBLP:conf/socs/StrasserHB14} is used as CPD.
For every distance estimation, a shortest path to the target is computed, whose length is used as the heuristic value.
Unfortunately, the employed CPD's quadratic preprocessing running time is problematic on large street networks.
In~\cite{DBLP:conf/ijcai/0002UJAKK18} the weighted graph is embedded into Euclidean space using FastMap. 
The Euclidean distance is then used as a distance estimate for A*.

\section{Algorithm}\label{sec:main-algo}

In this section, we first discuss the framework in which CH-Potentials can be used.
Then, we describe the building blocks of CH-Potentials: Contraction Hierarchies and PHAST, a CH extension.
Finally, we introduce the CH-Potentials heuristic.

\subsection{Formal Setup: Inputs, Outputs, and Phases}

In this paper, we consider different applications, with slightly different problem models.
The goal is always to quickly answer many shortest path queries.
For the purpose of describing our framework, we establish a shared notation:
Input to each query are nodes $s$ and $t$, and a graph $G_q$ with query weights $w_q$.
However, the precise formal inputs of the query and what exactly $w_q$ represents depends on the application.
In the simplest case, $w_q$ will be scalar edge weights.
However, this is not a requirement.
It can be any function that computes a weight for an edge.
This function can also take additional parameters from the state of the search.
For example, in the case of live-traffic, $w_q$ represents scalar edge weights.
However, values of $w_q$ might change between queries.
In the case of traffic predictions, $w_q$ is a function which maps the edge entry time to the traversal time and the query takes an additional departure time parameter.

To enable quick shortest path computations, we consider a two phase setup with an additional off-line preprocessing phase before the on-line query phase.
The input to the preprocessing phase is a graph $G_\ell$ with lower bound weights $w_\ell$ and a node mapping function $\phi$.
$w_\ell(e)$ must be a scalar value for every edge $e$ of $G_\ell$.
We require that $w_q(u,v)$ is greater or equal to the shortest distance $\operatorname{dist}_\ell(\phi(u), \phi(v))$ from $\phi(u)$ to $\phi(v)$ in $G_\ell$. 
The output of the preprocessing is auxiliary data that enables an efficient heuristic function $h_t(x)$.
$h_t(x)$ is the exact distance from $\phi(x)$ to $\phi(t)$ in $G_\ell$.
In the applications considered in this paper, $w_\ell$ is always the freeflow travel time.

The query phase uses this heuristic in an A* search between nodes $s$ and $t$ on $G_q$ and $w_q$.
The exact implementation of this A* search depends on the application.
Our approach only provides the heuristic $h_t$ for the A* search.
In contrast, the preprocessing phase remains the same for all applications.

Our heuristic is always \emph{feasible}~\cite{hnr-afbhd-68}, i.e. $w_q(u,v) - h_t(u) + h_t(v) \geq 0$ holds for all edges.
By requirement and because of the triangle inequality the following holds:
\[
w_q(u,v) - h_t(u) + h_t(v) \geq \operatorname{dist}_\ell(\phi(u), \phi(v)) - \operatorname{dist}_\ell(\phi(u), \phi(t)) + \operatorname{dist}_\ell(\phi(v), \phi(t)) \geq 0
\]
Thus, A* will always determine the correct shortest distances.

\subsection{Contraction Hierarchy (CH)}

\begin{algorithm2e}
\KwData{$B[x]$: tentative distance from $x$ to target $t$}
\KwData{Min. priority queue $Q$, also called open list}
$B[x] \leftarrow +\infty$ for all $x\neq t$;
$B[t] \leftarrow 0$\;
Make $Q$ only contain $t$ with weight $0$\;
\While{not $Q$ empty}{
	$y\leftarrow$ pop minimum element from $Q$\;
	\For{$xy$ is down-edge in $G^+_\ell$}{
		\If{$B[x] > w_\ell(xy) + B[y]$}{
			$B[x]\leftarrow w_\ell(xy) + B[y]$\;
                        Add $x$ or decrease $x$'s key in $Q$ to $B[x]$\;
		}
	}
}
\caption{CH backward search}
\label{algo:ch-backward}
\end{algorithm2e}

\begin{figure}
\centering
\includegraphics{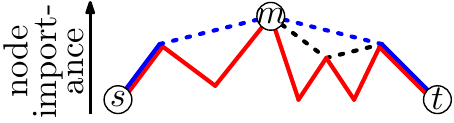}
\caption{
Solid lines are edges in $G$. Dotted lines are shortcuts. Red is shortest $st$-path in $G$. Blue is equaly long up-down $st$-path in $G^+$. $m$ is the mid node.
}
\label{fig:ch}
\end{figure}

A CH is a two phase technique to efficiently compute exact, shortest paths.
For details, we refer to \cite{gssv-erlrn-12,dsw-cch-15}.
In this section, we give an introduction.

A CH places nodes into levels.
No edge must connect two nodes within one level.
Levels are ordered by ``importance''.
The intuition is that dead-ends are unimportant and at the bottom while highway bridges are very important and at the top.
An edge goes \emph{up} when it goes from a node in a lower level to higher level.
\emph{Down} edges are defined analogously.
An \emph{up-down path} is a path where only one node $m$ is more important than both its neighbors.
$m$ is called the \emph{mid} node.
An \emph{up path} is a path where the last node is the mid node.
Similarly, the first node is the mid node of a \emph{down path}.
In the preprocessing phase, a CH adds \emph{shortcut} edges to the input graph $G$ to obtain $G^+$.
This is done by repeatedly contracting unimportant nodes and adding shortcuts between its neighbors.
After the preprocessing, for every pair of nodes $s$ and $t$ there exists a shortest up-down $st$-path in $G^+$ with the same length as a shortest path in $G$.
See Figure~\ref{fig:ch} for a proof sketch.
From every shortest path (red) in $G$, an up-down path of equal length in $G+$ (blue) exists.
Thus, we can restrict our search to up-down paths in $G^+$.
The search is bidirectional.
The forward search starts from $s$ and only follows up-edges.
Similarly, the backward search starts at $t$ and only follows down-edges in reversed direction.
The two searches meet at the mid node.
Pseudo-code for the backward search, i.e., the path from $m$ to $t$, is presented in Algorithm~\ref{algo:ch-backward}.
The forward search works analogously.
A CH query is fast, if the number of nodes reachable via only up- or down-nodes is small.
On road networks, this is the case~\cite{gssv-erlrn-12,dgpw-crprn-13}.
On graphs with low treewidth, this is also the case~\cite{dsw-cch-15,hs-gbpo-18}.

Using the CH query algorithm, we can already give a simple heuristic.
The heuristic evaluation $h_t(x)$ performs a CH-query from $x$ to $t$.
This yields tight estimates but a high overhead for the heuristic evaluation.
While a single CH query is fast, answering one for every node explored in the $A^*$ search is slow.
Fortunately, we can do better.

\subsection{PHAST based Heuristic}

\begin{algorithm2e}
\KwData{$P[x]$: tentative distance from $x$ to $t$}
Execute Algorithm~\ref{algo:ch-backward}\;
\For{all CH levels $L$ from most to least important}{
	\For{all up edges $xy$ in $G^+_\ell$ with $x$ in $L$}{
		\If{$P[x] < P[y] + w_\ell(xy)$}{
			$P[x] \leftarrow P[y] + w_\ell(xy)$\;
		}
	}
}
\caption{PHAST basic all-to-one search}
\label{algo:phast}
\end{algorithm2e}

PHAST~\cite{dgnw-phast-13} is a CH extension that computes distances from all nodes to one target node.
The preprocessing phase remains unchanged.
The query phase is split into two steps.
The first step is analogue to the CH query:
From $t$, all reachable nodes via reversed down-edges are explored.
Algorithm~\ref{algo:ch-backward} shows this first step.
The second step iterates over all CH levels from top to bottom.
In each iteration, all up-edges starting within the current level are followed in reverse.
After all levels are processed, the shortest distances from all nodes to $t$ were computed.
Pseudo-code is provided in Algorithm~\ref{algo:phast}.
Using PHAST, we can also compute a tight A* heuristic.
In the query phase, we first run PHAST to compute the distances from every node to $t$ with respect to $w_\ell$ and store the result in an array $H$.
Next, we run $A^*$ and implement the heuristic as a lookup in the array $H$.

The $H$ lookup and by extension the $A^*$ search is indeed fast.
However, the PHAST step before the search is comparatively expensive.
The reason is that the distances towards $t$ are computed for \emph{all} nodes.
Ideally, we only want to compute the distances from the nodes explored in the $A^*$ search.

\subsection{CH-Potentials}

\begin{algorithm2e}
\KwData{$B[x]$: tentative distance from $x$ to $t$ as computed by Algorithm~\ref{algo:ch-backward}}
\KwData{$P[x]$: memoized potential at $x$, $\bot$ initially}
\SetKwFunction{Pot}{Pot}
\SetKwProg{Fn}{Function}{:}{}
\Fn{\Pot{$x$}}{
	\If{$P[x] = \bot$}{
		$P[x]\leftarrow B[x]$\;
		\For{all up edges $xy$ in $G^+_\ell$}{
                        $P[x]\leftarrow\min\{P[x],w_\ell(xy)+\Pot(y)\}$\;
		}
	}
	\Return{$P[x]$}\;
}
\caption{CH-Potentials Algorithm}
\label{algo:pot}
\end{algorithm2e}

Fortunately, the PHAST computation can be done lazily using memoization as depicted in Algorithm~\ref{algo:pot}.
In a first step, we run the backward CH search from $t$ to obtain an array $B$.
$B[x]$ is the minimum down $xt$-path distance or $+\infty$, if there is no such path.
$B$ is computed as shown in Algorithm~\ref{algo:ch-backward}.

To compute the heuristic $h_t(x)$, we recursively compute for all up-edges $(x,y)$ the heuristic $h_t(y)$.
Next, we compute the minimum distance over all up-down paths that contain at least one up-edge using $d = \min_y\{w_\ell(x,y) + h_t(y)\}$.
As not all shortest up-down paths contain an up-edge, we set $h_t(x) = \min \{ B[x], d \}$.
This calculation is correct, as it computes the minimum up-down $xt$-path distance in $G^+_\ell$, which corresponds to the minimum $xt$-path distance in a CH.
A* with this heuristic is the basic CH-Potentials algorithm.

\section{Low Degree A* Improvements}

\label{sec:low-deg-improvment}

Preliminary experiments showed, that a significant amount of query running time is spent in heuristic evaluations and queue operations.
We can reduce both by keeping some nodes out of the queue, as the heuristic needs to be evaluated when a node is pushed into the queue.
Avoiding pushing low degree nodes into the queue is the focus of this section.
The techniques discussed here are a lazy variant of the ideas used in TopoCore~\cite{DBLP:conf/gis/DibbeltSW15}.

We modify A* by processing low degree nodes consecutively without pushing them into the queue.
Our algorithm uses the undirected degree $d(x)$ of a node $x$.
Formally, $d(x)$ is the number of nodes $y$ such that $(x,y)\in E$ or $(y,x)\in E$.

Analogous to A*, our algorithm stores for every node $x$ a tentative distance $D[x]$.
Additionally, it maintains a minimum priority queue.
Diverging from A*, not all nodes can be pushed but every node has a tentative distance.

\subsection{Skip Degree Two Nodes}

Our algorithm differs from A* when removing a node $x$ from the queue.
A* iterates over the outgoing arcs $(x,y)$ of $x$ and tries to reduce $D[y]$ by relaxing $(x,y)$.
If A* succeeds, $y$'s weight in the queue is set to $D[y]+h_t(y)$.
Our algorithm, however, behaves differently, if $d(y)\le 2$.
Our algorithm determines the longest degree two chain of nodes $x,y_1,\ldots, y_k, z$ such that $d(y_i)=2$ and $d(z) > 2$.
If our algorithm succeeds in reducing $D[y_1]$, it does not push $y_1$ into the queue.
Instead, it iteratively tries to reduce all $D[y_i]$.
If it does not reach $z$, then only $D$ is modified but no queue action is performed.
If $D[z]$ is modified and $d(z)>2$, $z$'s weight in the queue is set to $D[z]+h_t(z)$.

As the target node $t$ might have degree two, our algorithm cannot rely on stopping, when $t$ is removed from the queue.
Instead, our algorithm stops as soon as $D[t]$ is less than the minimum weight in the queue.

\subsection{Skip Degree Three Nodes}

We can also skip some degree three nodes.
Denote by $x,y_1,\ldots, y_k, z$ a degree two chain as described in the previous section.
If $d(z) > 3$ or $z$ is in the queue, our algorithm proceeds as in the previous section.
Otherwise, there exist up to two degree chains $z,a_1,\ldots,a_p,b$ and $z,\alpha_1,\ldots,\alpha_q,\beta$ such that $a_1\neq y_k \neq \alpha_1$.
Our algorithm iteratively tries to reduce all $D[a_i]$ and $D[\alpha_i]$.
If it reaches $\beta$, $\beta$'s weight in the queue is set to $D[\beta]+h_t(\beta)$.
Analogously, if $b$ is reached, $b$'s weight is set to $D[b]+h_t(b)$.
If $b$ respectively $\beta$ are not reached, our algorithm does nothing.

\subsection{Stay in Largest Biconnected Component}
\label{sec:largested-biconnected-component}

A lot of nodes in road networks lead to dead-ends.
Unless the source or target is in this dead-end, it is unnecessary to explore these nodes.

In the preprocessing phase, we compute the subgraph $G_C$, called \emph{core}.
$G_C$ is induced by the largest biconnected component of the undirected graph underlying $G$.
We do this using Tarjan's algorithm \cite{t-dfslg2-72}.
For every node $v$ in the input graph $G$, we store the attachment node $a_v$ to the core.
For nodes in the core, $a_v=v$.
We exploit that all attachment nodes are single node separators and the problem can be decomposed along them.

The query phase is divided into two steps.
In the first step, we apply A* with CH-Potentials to $G_C$ combined with the component that contains $s$.
This can be achieved implicitly by removing edges from $G_C$ into other components during preprocessing.
If $t$ is part of $G_C$ or in the same component as $s$, this A* search finds it.
Otherwise, we find $a_t$.
In that case, we continue by searching a path from $a_t$ towards $t$ restricted to $t$'s biconnected component.
The final result is the concatenation of both paths.

\section{Applications}
\label{sec:extensions}

We describe some extended routing problems and how to apply CH-Potentials to them.
Unless stated otherwise, $G_q$ and $G_\ell$ are the same graph and only $w_q$ changes for the queries.

\subsection{Avoiding Tunnels and/or Highways}
\label{sec:no-tunnel-highway}

Avoiding tunnels and/or highways is a common feature of navigation devices.
Implementing this feature with CH-Potentials is easy.
We set $w_\ell$ to the freeflow travel time.
If an edge is a tunnel and/or a highway, we set $w_q$ to $+\infty$.
Otherwise, $w_q$ is set to the freeflow travel time.

\subsection{Forbidden Turns and Turn Costs}
\label{sec:no-turns}

The classical shortest path problem allows to freely change edges at nodes.
However, in the real world, turn restrictions, such as a forbidden left or right turn, exist.
Also, taking a left turn might take longer than going straight.
This can be modeled using turn weights~\cite{gv-errnt-11,dgpw-crprn-13,bwzz-cchtc-20}.
A \emph{turn weight} $w_t$ maps a pair of incident edges onto the turning time or $+\infty$ for forbidden turns.
For CH-Potentials, we use zero as lower bound for every turn weight in the heuristic.
Thus, the graph $G_\ell$ and weights $w_\ell$ for preprocessing is the unmodified input graph without turn weights.

A path with nodes $v_1, v_2,\ldots v_k$ has the following \emph{turn-aware weight}: \[
w_\ell(v_1, v_2) + \sum_{i=2}^{k-1}  w_t(v_{i-1},v_i,v_{i+1}) + w_\ell(v_i,v_{i+1})
\]
The objective is to find a path between two given edges with minimum turn-aware weight.
The first term $w_\ell(v_1, v_2)$ is the same for all paths, as it only depends on the source edge.
It can thus be ignored during optimization.

We solve this problem by constructing a \emph{turn-expanded} graph as $G_q$.
Edges in the input graph $G_\ell$ correspond to \emph{expanded nodes} in $G_q$.
For every pair of incident edges $(x,y)$ and $(y,z)$ in $G_\ell$, there is an \emph{expanded edge} in $G_q$ with expanded weight $w_t(x,y,z) + w_\ell(y,z)$.
A sequence of expanded nodes in the expanded graph $G_q$ corresponds to a sequence of edges in the input graph $G_\ell$.
The weight of a path in $G_q$ is equal to the turn-aware weight of the corresponding path in $G_\ell$ minus the irrelevant $w_\ell(v_1,v_2)$ term.
Thus, the turn-aware routing problem can be solved by searching for shortest paths in $G_q$.

In this scenario, preprocessing and query use different graphs $G_\ell$ and $G_q$.
We define the node mapping function $\phi$ as $\phi(x,y) = y$.
Obviously, $w_q(xy, yz) = w_t(x,y,z) + w_\ell(y,z) \geq \operatorname{dist}_\ell(\phi(x,y), \phi(y,z))$ and this approach yields a feasible heuristic.
Sadly, the undirected graph underlying $G_q$ is always biconnected, if the input graph is strongly connected.
The optimization described in Section~\ref{sec:largested-biconnected-component} is therefore ineffective.
With this setup, CH-Potentials support turns without requiring turn information in the CH.

\subsection{Predicted Traffic or Time-Dependent Routing}
\label{sec:predicted-traffic}

The classical shortest path problem assumes that edge weights are scalars.
However, in practice, travel times vary along an edge due to the traffic situation.
Recurring traffic can be predicted by observing the traffic in the past.
It is common \cite{bgsv-mtdtt-13,bdpw-dtdrp-16,swz-sfert-20} to represent these predictions as \emph{travel time functions}.
An edge weight is no longer a scalar value but a function that maps the entry time onto the traversal time.

In this setting, the query weight $w_q$ is a function from $E\times \mathbb{R}$ to $\mathbb{R}^+$.
$w_q(e, \tau)$ is the travel time through edge $e$ when entering it at moment $\tau$.
The input to the extended problem consists of a source node $s$ and a target node $t$, as in the classical problem formulation.
Additionally, the input contains a source time $\tau_s$.
A path with edges $e_1,e_2\ldots e_k$ is weighted using $\alpha_k$, which is defined recursively as follows:\[
\begin{split}
\alpha_{1} & = w_q(e_1, \tau_s) \\
\alpha_{k} & = \alpha_{k-1} + w_q(e_1, \alpha_{k-1})
\end{split}
\]
The objective is to find a path to $t$ that minimizes $\alpha_k$.

If all travel time functions fulfill the \emph{FIFO property}, this problem can be solved using a straight forward extension of Dijkstra's algorithm \cite{d-aassp-69}.
The necessary modification to A* is analogous.
Without the FIFO property the problem becomes NP-hard \cite{or-tnp-89}.
The FIFO property states that it is not possible to arrive earlier by departing later.
Formally stated, the following must hold $\forall e\in E,\tau\in \mathbb{R},\delta\in \mathbb{R}^+: w_q(e, \tau) \le w_q(e, \tau+\delta) + \delta$.
Our implementation stores edge travel times using piece-wise linear functions.
The A* search uses the tentative distance $\tau$ at a node $x$ when to evaluating the travel time of outgoing edges $(x,y)$.
This strategy is very similar to TD-ALT~\cite{ndls-bastd-12,dw-lbrdg-07}.

For the preprocessing, we set $w_\ell(e) = \min_\tau w_q(e,\tau)$, that is the minimum travel time.
By keeping travel time functions out of the CH, we avoid a lot of algorithmic complications compared to~\cite{bgsv-mtdtt-13,bdpw-dtdrp-16,swz-sfert-20,dn-crdtd-12} which have to create shortcuts of travel time functions.

\subsection{Live and Predicted Traffic}
\label{sec:live-predicted-traffic}

Beside predicted traffic, we also consider live traffic.
Live traffic refers to the current traffic situation.
It is important to distinguish between predicted and live traffic.
Live traffic data is more accurate for the current moment than predicted data.
It is possible that it differs significantly from predicted traffic, if unexpected events like accidents happen.
However, just using live traffic data is problematic for long routes as traffic changes while driving.
At some point, one wants to switch from live traffic to the predicted traffic.
In this section, we first describe a setup with only live traffic and then combine it with predicted traffic.

To support only live traffic, we set $w_\ell$ to the freeflow travel time.
$w_q$ is set to the travel time accounting for current traffic.
As traffic only increases the travel time along an edge, $w_\ell$ is a valid lower bound for $w_q$.
In a real world application, values from $w_q$ could be updated between queries.
This is all that is necessary to apply CH-Potentials in a live traffic scenario.

To combine live traffic with predicted traffic, we define a modified travel time function $w_q$ that is then used as query weights.
Denote by $w_p(e,\tau)$ the predicted travel time along edge $e$ at moment $\tau$.
Further, $w_c(e)$ is the travel time according to current live traffic.
Finally, we denote by $\tau_{\mathrm{soon}}$ the moment when we switch to predicted traffic.
In our experiments, we set $\tau_{\mathrm{soon}}$ to one hour in the future.
We need to make sure that the modified travel time function fulfills the no-waiting property.
For this reason, we cannot make a hard switch at $\tau_{\mathrm{soon}}$.
Our modified travel time function linearly approaches the predicted travel time. 
Formally, we set $w_q(e,\tau)$ to $w_c(e)$, if $\tau \leq \tau_{\mathrm{soon}}$.
Otherwise, we check whether $w_p(e,\tau_{\mathrm{soon}}) < w_c(e)$ is true.
If it is the case, we set $w_q(e,\tau)$ to $\max\{w_c(e)+(\tau_{\mathrm{soon}}-\tau), w_p(e,\tau)\}$.
Otherwise, we set $w_q(e,\tau)$ to $\min\{w_c(e)-(\tau_{\mathrm{soon}}-\tau), w_p(e,\tau)\}$.
In our implementation, we to not modify the representation of $w_p$ but evaluate the formulas above at each travel time evaluation.
We set $w_\ell$ again to the freeflow travel time.

With this setup, CH-Potentials support a combination of live and predicted traffic.
We did not make any modification, that would hinder a combination with other extensions.
Further adding tunnel and/or highway avoidance or turn-aware routing is simple.
This straight-forward integration of complex routing problems is the strength of the CH-Potentials.

\subsubsection{Three-Phase Setups}

Supporting live traffic is also possible with a three-phase setup:
A slow preprocessing phase, a faster \emph{customization} phase, and fast queries.
The customization phase is run regularly and incorporates updates to the weights into the auxiliary preprocessing data.
CRP~\cite{dgpw-crprn-13} and CCH~\cite{dsw-cch-15} follow this setup.
Luckily, a CCH is just a CH with some additional properties.
The CH in CH-Potentials can be replaced by a CCH without further modifications.
Thus, CCH-Potentials could also support a three-phase setup.
However, evaluating CCH-Potentials is beyond the scope of this paper.
We focus on evaluating CH-Potentials as a simple building block in the two-phase setup.

\subsection{Temporary Driving Bans}

Truck routing differs from car routing due to night driving bans and other restrictions.
In~\cite{kswz-erptd-p-20}, a preliminary version of CH-Potentials\footnote{
In~\cite{kswz-erptd-p-20}, CH-Potentials are used as a blackbox referring to an early ArXiv preprint~\cite{strasser2019perfect} of ours.
Our submitted paper is the finished version of the ArXiv preprint.
\cite{kswz-erptd-p-20} does not describe any of the contributions of this paper.
} is used for such a scenario.
The work considers time-dependent blocked edges and waiting at parking locations.
Further, a trade-off between arrival time and route quality is considered.

\section{Evaluation}

\label{sec:experiments}

\begin{table}
\centering
\caption{Instances used in the evaluation.}\label{tab:graphs}
\begin{tabular}{lrrr}
\toprule
 &          Nodes &          Edges & Preprocessing \\ & $[\cdot 10^6]$ & $[\cdot 10^6]$ &           [s] \\
\midrule
OSM Ger &       16.2 &       35.4 &                         295.2 \\
TDEur17 &       25.8 &       55.5 &                         292.6 \\
TDGer06 &        4.7 &       10.8 &                          58.9 \\
\bottomrule
\end{tabular}

\end{table}

In this section, we present our experimental evaluation.
Our benchmark machine runs openSUSE Leap 15.1 (kernel 4.12.14), and has 128\,GiB of DDR4-2133 RAM and an Intel Xeon E5-1630 v3 CPUs which has four cores clocked at 3.7\,Ghz and 4~$\times$~32\,KiB of L1, 8~$\times$~256\,KiB of L2, and 10\,MiB of shared L3 cache.
All experiments were performed sequentially.
Our code is written in Rust and compiled with rustc 1.47.0-nightly in the release profile with the target-cpu=native option.
The source code of our implementation and the experimental evaluation can be found on Github\footnote{\url{https://github.com/kit-algo/ch_potentials}}.

\subparagraph{Inputs and Methodology}
Our main benchmark instance is a graph of the road network of Germany obtained from Open Street Map\footnote{\url{https://download.geofabrik.de/europe/germany-200101.osm.pbf}}.
To obtain the routing graph, we use the import from RoutingKit\footnote{\url{https://github.com/RoutingKit/RoutingKit}}.
The graph has 16M nodes and 35M edges.
For this instance, we have proprietary traffic data provided by Mapbox\footnote{\url{https://mapbox.com}}.
The data includes a live traffic snapshot from Friday 2019/08/02 afternoon and comes in the form of 320K OSM node pairs and live speeds for the edge between the nodes.
It also includes traffic predictions for 38\% of the edges as predicted speeds for all five minute periods over the course of a week.
We exclude speed values which are faster than the freeflow speed computed by RoutingKit.
Additionally, we have two graphs with proprietary traffic predictions provided by PTV\footnote{\url{https://ptvgroup.com}}.
The PTV instances are not OSM-based.
One is an old instance of Germany with traffic predictions from 2006 for 7\% of the edges and the other one a newer instance of Europe with predictions for 27\% of the edges.
Table~\ref{tab:graphs} contains an overview over our instances.
In this table, we further include the sequential running time necessary to construct the CH.
We report preprocessing running times as averages over 10 runs.
For queries, we perform 10\,000 point-to-point queries where both source and target are nodes drawn uniformly at random and report average results.

\subparagraph{Experiments}

\begin{figure}
\centering
\includegraphics[width=\linewidth]{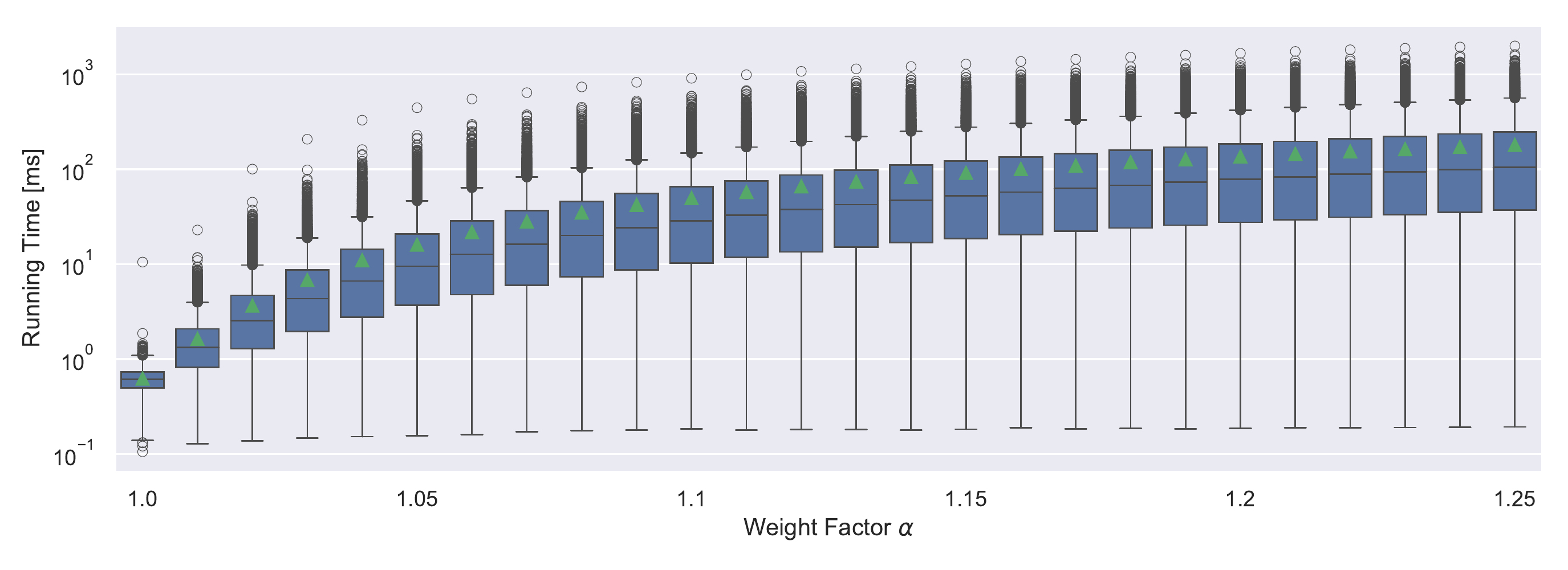}
\caption{
Running times on a logarithmic scale for queries on OSM Ger with scaled edge weights $w_q = \alpha \cdot w_\ell$.
The boxes cover the range between the first and third quartile.
The band in the box indicates the median, the diamond the mean.
The whiskers cover 1.5 times the interquartile range.
All other running times are indicated as outliers.
}\label{fig:scaled_weights}
\end{figure}

The performance of A* depends on the tightness of the heuristic.
CH-Potentials computes optimal distance estimates with respect to $w_\ell$.
However, for most applications, there will be a gap between $w_q$ and $w_\ell$ (otherwise one could use CH without A*).
We evaluate the impact of the difference between $w_q$ and $w_\ell$ on the performance of A*.
The lower bound $w_\ell$ is set to the freeflow travel time.
The query weights $w_q$ are set to $\alpha \cdot w_\ell$, where $\alpha\ge 1$.
Increasing $\alpha$ degrades the heuristic's quality.
Figure~\ref{fig:scaled_weights} depicts the results.
Clearly, $\alpha$ has significant influence on the running time.
Average running times range from below a millisecond to a few hundred milliseconds depending on $\alpha$.
Up to around $\alpha = 1.1$ the running time grows quickly.
For $\alpha > 1.1$, the growth slows down.
This illustrates both the strengths and limits of our approach and goal directed search in general.
CH-Potentials can only achieve competitive running times if the application allows for a sufficiently tight lower bounds at preprocessing time.

We observe that the running times for a fixed $\alpha$ vary strongly.
This is an interesting observation, as with uniform source and target sampling, nearly all queries are long-distance.
The query distance is thus not the reason.
After some investigation, we concluded that this is due to non-uniform road graph density.
Some regions have more roads per area than others.
The number explored A* nodes depends on the density of the search space area.
As the density varies, the running times vary.

\begin{table}
\centering
\caption{Average query running times and number of queue pushs with different heuristics and optimizations on OSM Ger with $w_q = 1.05 \cdot w_\ell$.}\label{tab:building_blocks}
\begin{tabular}{clllrrrr}
\toprule
       & BCC & Deg2 & Deg3 & Zero & ALT & CH-Pot. & Oracle \\
\midrule
\multirow{4}{*}{\rotatebox[origin=c]{90}{\shortstack{Running\\time [ms]}}} & \xmark &        \xmark &        \xmark &  1\,947.4 &  279.5 &   50.6 &    32.0 \\
                                                                                    & \cmark  &        \xmark &        \xmark &  1\,253.1 &  217.3 &   36.3 &    23.9 \\
                                                                                    & \cmark  &         \cmark &        \xmark &   713.6 &  117.3 &   18.8 &    11.8 \\
                                                                                    & \cmark  &         \cmark &         \cmark &   558.9 &   88.3 &   15.7 &     9.5 \\
\multirow{4}{*}{\rotatebox[origin=c]{90}{\shortstack{Queue\\pushs [$\cdot 10^3$]}}} & \xmark &        \xmark &        \xmark &  8\,120.7 &  859.4 &  138.0 &   138.0 \\
                                                                                    & \cmark  &        \xmark &        \xmark &  6\,326.5 &  684.1 &  114.0 &   114.0 \\
                                                                                    & \cmark  &         \cmark &        \xmark &  2\,915.7 &  301.3 &   42.1 &    42.1 \\
                                                                                    & \cmark  &         \cmark &         \cmark &  1\,689.8 &  178.4 &   26.0 &    26.0 \\
\bottomrule
\end{tabular}

\end{table}

Table~\ref{tab:building_blocks} depicts the performance of A* with different heuristics and optimizations.
We compare CH-Potentials to three other heuristics.
First, the Zero heuristic where $h(x)=0$ for all nodes $x$.
This corresponds to using Dijkstra's algorithm.
Second, we compare against our own implementation of ALT~\cite{gw-cppsp-05}.
We use 16 landmarks generated with the avoid strategy~\cite{gw-cppsp-05} and activate all during every query.
Our ALT implementation is uni-directional.
In this work, we do not consider bidirectional search as it creates problems for some settings, such as predicted traffic.
Finally, we compare against a hypothetical \emph{Oracle-A*} heuristic.
This heuristic has instant access to a shortest distance array with respect to $w_\ell$, i.e. it is faster than the fastest heuristic possible in our model.
We fill this array before each query using a reverse Dijkstra search from the target node.
Thus, the reported running times of Oracle-A* do \emph{not} account for any heuristic evaluation.
CH-Potentials compute the same distance estimates but the heuristic evaluation has some overhead.
Comparing against Oracle-A* allows us to measure this overhead.
Also, no other heuristic, which only has access to the preprocessing weights, can be faster than Oracle-A*.

We observe that the number of queue pushes roughly correlates with running time.
Each optimization reduces both queue pushes and running times.
All optimizations yield a combined speed-up of around 3.
CH-Potentials outperform ALT by a factor of between six and seven and settle correspondingly fewer nodes.
This is not surprising, since ALT computes worse distance estimates.
In contrast, CH-Potentials already compute exact distances with respect to $w_\ell$.
The number of popped nodes is the same for CH-Potentials and Oracle-A*.
The only difference between CH-Potentials and Oracle-A* is the overhead of the heuristic evaluation.
This overhead leads to a slowdown of around 1.6.
Thus, CH-Potentials are already very close to the best possible heuristic in this model.
This means that no competing algorithm such as ALT or CPD-Heuristics can be significantly faster.

\begin{table*}
\centering
\caption{
CH-Potentials performance for different route planning applications.
We report average running times and number of queue pushes.
We also report the average length increase, that is how much longer the final shortest distance is compared to the lower bound.
Finally, we report the average running time of Dijkstra's algorithm as a baseline and the speedup over this baseline.
}\label{tab:applications}
\begin{tabular}{llrrrrr}
\toprule
 & &   Running &                Queue &     Length & Dijkstra & Speedup \\ & & time [ms] & $[\cdot 10^3]$ & incr. [\%] &     [ms] &         \\
\midrule
\multirow{8}{*}{OSM Ger} & Unmodified ($w_q=w_\ell$) &              0.6 &              0.5 &       0.0 &                    1\,952.8 &   3\,243.1 \\
        & Turns &              2.8 &              6.0 &       1.1 &                    4\,244.4 &   1\,540.8 \\
        & No Tunnels &             25.9 &             40.7 &       5.3 &                    1\,990.1 &     76.9 \\
        & No Highways &            342.9 &            518.5 &      42.4 &                    1\,843.9 &      5.4 \\
        & Live &            127.3 &            192.1 &      14.8 &                    1\,884.1 &     14.8 \\
        & TD &            195.5 &            163.1 &      17.6 &                    3\,186.2 &     16.3 \\
        & TD + Live &            209.7 &            179.1 &      21.4 &                    3\,152.2 &     15.0 \\
        & TD + Live + Turns &            508.5 &            765.5 &      22.7 &                    6\,179.5 &     12.2 \\
\addlinespace
TDEur17 & TD &             89.7 &             81.5 &       3.9 &                    3\,479.9 &     38.8 \\
TDGer06 & TD &              4.5 &              6.4 &       3.1 &                     602.4 &    135.4 \\
\bottomrule
\end{tabular}

\end{table*}

Table~\ref{tab:applications} depicts the running times of CH-Potentials in various applications, such as those described in Section~\ref{sec:extensions}.
We report speedups compared to extensions of Dijkstra's algorithm for each application respectively.
We start with the base case where $w_q = w_\ell$.
This is the problem variant solved by the basic CH algorithm.
CH achieves average query running times of 0.16\,ms on OSM Ger.
CH-Potentials are roughly four times slower but still achieve a huge speedup of 3243 over Dijkstra.
Such large speedups are typical for CH.
This shows that CH-Potentials gracefully converges toward a CH in the $w_q = w_\ell$ special case.

In the other scenarios, the performance of CH-Potentials strongly depends on the quality of the heuristic.
We measure this quality using the length increase of $w_q$ compared to $w_\ell$.
Forbidding highways results in the largest length increase and in the smallest speedup.
The other extreme are turn restrictions.
They have only a small impact on the length increase.
The achieved speedups are therefore comparable to CH speedups.
Mapbox live traffic has a length increase of around 15\%, which yields running times of 127\,ms.
The length increase of Mapbox traffic predictions are about 18\%, and results in a running time of 200\,ms.
The speedup in the predicted scenario is larger than in the live setting, as the travel time function evaluations slow down Dijkstra's algorithm.
Combining predicted and live traffic results in a running time only slightly higher than for the predicted scenario.
Further adding turn restrictions, increases the running times.
This increase is mostly due to the BCC optimization of Section~\ref{sec:largested-biconnected-component} becoming ineffective when considering turns.
It is not due to the length increase of using turns.
With everything activated, our algorithm still has a speedup of 12.2 over the baseline.
Interestingly, the PTV traffic predictions have a much smaller length increase than the Mapbox predictions.
This results in smaller running times of our algorithm.

\section{Conclusion}
\label{sec:conclusion}

In this paper, we introduced CH-Potentials, a fast, exact, and flexible two-phase algorithm based on A* and CH for finding shortest paths in road networks.
The approach can handle a multitude of complex, integrated routing scenarios with very little implementation complexity.
CH-Potentials provides \emph{exact} distances with respect to lower bound weights known at preprocessing time as an A* heuristic.
Thus, the query performance of CH-Potentials crucially depends on the availability of good lower bounds in the preprocessing phase.
Our experiments show, that this availability highly depends on the application.
We also show that the overhead of our heuristic is within a factor 1.6 of a hypothetical A*-heuristic that can instantly access lower bound distances.
Achieving significantly faster running times could still be possible in variations of the problem setting.

Dropping the provable exactness requirement using a setup similar to anytime A*~\cite{DBLP:conf/aaai/ZhouH02,DBLP:conf/nips/LikhachevGT03} would be interesting.
Another promising research avenue would be to investigate graphs other than road networks.
A lot of research into grid maps exists including a series of competitions called GPPC~\cite{DBLP:conf/socs/SturtevantTTUKS15}.
Hierarchical techniques have been shown to work well on these graphs~\cite{DBLP:conf/aaai/UrasK14}.


\begin{thebibliography}{10}

\bibitem{bdgmpsww-rptn-16}
Hannah Bast, Daniel Delling, Andrew~V. Goldberg, Matthias
  {M{\"u}ller--Hannemann}, Thomas Pajor, Peter Sanders, Dorothea Wagner, and
  Renato~F. Werneck.
\newblock {Route Planning in Transportation Networks}.
\newblock In Lasse Kliemann and Peter Sanders, editors, {\em Algorithm
  Engineering - Selected Results and Surveys}, volume 9220 of {\em Lecture
  Notes in Computer Science}, pages 19--80. Springer, 2016.

\bibitem{b-tdrpc-14}
Gernot~Veit Batz.
\newblock {\em {Time-Dependent Route Planning with Contraction Hierarchies}}.
\newblock PhD thesis, Karlsruhe Institute of Technology (KIT), 2014.

\bibitem{bdsv-tdch-09}
Gernot~Veit Batz, Daniel Delling, Peter Sanders, and Christian Vetter.
\newblock {Time-Dependent Contraction Hierarchies}.
\newblock In {\em Proceedings of the 11th Workshop on Algorithm Engineering and
  Experiments (ALENEX'09)}, pages 97--105. SIAM, April 2009.

\bibitem{bgns-tdcha-10}
Gernot~Veit Batz, Robert Geisberger, Sabine Neubauer, and Peter Sanders.
\newblock {Time-Dependent Contraction Hierarchies and Approximation}.
\newblock In Paola Festa, editor, {\em Proceedings of the 9th International
  Symposium on Experimental Algorithms (SEA'10)}, volume 6049 of {\em Lecture
  Notes in Computer Science}, pages 166--177. Springer, May 2010.
\newblock URL: \url{http://www.springerlink.com/content/u787292691813526/}.

\bibitem{bgsv-mtdtt-13}
Gernot~Veit Batz, Robert Geisberger, Peter Sanders, and Christian Vetter.
\newblock {Minimum Time-Dependent Travel Times with Contraction Hierarchies}.
\newblock {\em ACM Journal of Experimental Algorithmics}, 18(1.4):1--43, April
  2013.

\bibitem{bdsssw-chgds-10}
Reinhard Bauer, Daniel Delling, Peter Sanders, Dennis Schieferdecker, Dominik
  Schultes, and Dorothea Wagner.
\newblock {Combining Hierarchical and Goal-Directed Speed-Up Techniques for
  {D}ijkstra's Algorithm}.
\newblock {\em ACM Journal of Experimental Algorithmics}, 15(2.3):1--31,
  January 2010.
\newblock Special Section devoted to WEA'08.

\bibitem{bdgwz-sfpcs-19}
Moritz Baum, Julian Dibbelt, Andreas Gemsa, Dorothea Wagner, and Tobias
  Z{\"u}ndorf.
\newblock {Shortest Feasible Paths with Charging Stops for Battery Electric
  Vehicles}.
\newblock {\em Transportation Science}, 2019.

\bibitem{DBLP:journals/algorithmica/BaumDPSWZ20}
Moritz Baum, Julian Dibbelt, Thomas Pajor, Jonas Sauer, Dorothea Wagner, and
  Tobias Z{\"{u}}ndorf.
\newblock Energy-optimal routes for battery electric vehicles.
\newblock {\em Algorithmica}, 82(5):1490--1546, 2020.
\newblock \href {https://doi.org/10.1007/s00453-019-00655-9}
  {\path{doi:10.1007/s00453-019-00655-9}}.

\bibitem{bdpw-dtdrp-16}
Moritz Baum, Julian Dibbelt, Thomas Pajor, and Dorothea Wagner.
\newblock {Dynamic Time-Dependent Route Planning in Road Networks with User
  Preferences}.
\newblock In {\em Proceedings of the 15th International Symposium on
  Experimental Algorithms (SEA'16)}, volume 9685 of {\em Lecture Notes in
  Computer Science}, pages 33--49. Springer, 2016.

\bibitem{DBLP:conf/ijcai/BonoGHS19}
Massimo Bono, Alfonso~Emilio Gerevini, Daniel~Damir Harabor, and Peter~J.
  Stuckey.
\newblock Path planning with {CPD} heuristics.
\newblock In Sarit Kraus, editor, {\em Proceedings of the Twenty-Eighth
  International Joint Conference on Artificial Intelligence, {IJCAI} 2019,
  Macao, China, August 10-16, 2019}, pages 1199--1205. ijcai.org, 2019.
\newblock \href {https://doi.org/10.24963/ijcai.2019/167}
  {\path{doi:10.24963/ijcai.2019/167}}.

\bibitem{bwzz-cchtc-20}
Valentin Buchhold, Dorothea Wagner, Tim Zeitz, and Michael Z{\"u}ndorf.
\newblock {Customizable Contraction Hierarchies with Turn Costs}.
\newblock In Dennis Huisman and Christos Zaroliagis, editors, {\em Proceedings
  of the 20th Symposium on Algorithmic Approaches for Transportation Modelling,
  Optimization, and Systems (ATMOS'20)}, OpenAccess Series in Informatics
  (OASIcs), 2020.
\newblock Accepted for publication.

\bibitem{DBLP:conf/ijcai/0002UJAKK18}
Liron Cohen, Tansel Uras, Shiva Jahangiri, Aliyah Arunasalam, Sven Koenig, and
  T.~K.~Satish Kumar.
\newblock The fastmap algorithm for shortest path computations.
\newblock In J{\'{e}}r{\^{o}}me Lang, editor, {\em Proceedings of the
  Twenty-Seventh International Joint Conference on Artificial Intelligence,
  {IJCAI} 2018, July 13-19, 2018, Stockholm, Sweden}, pages 1427--1433.
  ijcai.org, 2018.
\newblock \href {https://doi.org/10.24963/ijcai.2018/198}
  {\path{doi:10.24963/ijcai.2018/198}}.

\bibitem{dgnw-phast-13}
Daniel Delling, Andrew~V. Goldberg, Andreas Nowatzyk, and Renato~F. Werneck.
\newblock {{PHAST}: Hardware-accelerated shortest path trees}.
\newblock {\em Journal of Parallel and Distributed Computing}, 73(7):940--952,
  2013.

\bibitem{dgpw-crprn-13}
Daniel Delling, Andrew~V. Goldberg, Thomas Pajor, and Renato~F. Werneck.
\newblock {Customizable Route Planning in Road Networks}.
\newblock {\em Transportation Science}, 51(2):566--591, 2017.

\bibitem{dn-crdtd-12}
Daniel Delling and Giacomo Nannicini.
\newblock {Core Routing on Dynamic Time-Dependent Road Networks}.
\newblock {\em Informs Journal on Computing}, 24(2):187--201, 2012.

\bibitem{dss-tarrn-18}
Daniel Delling, Dennis Schieferdecker, and Christian Sommer.
\newblock {Traffic-Aware Routing in Road Networks}.
\newblock In {\em Proceedings of the 34rd International Conference on Data
  Engineering}. IEEE Computer Society, 2018.
\newblock URL: \url{https://doi.org/10.1109/ICDE.2018.00172}.

\bibitem{dw-lbrdg-07}
Daniel Delling and Dorothea Wagner.
\newblock {Landmark-Based Routing in Dynamic Graphs}.
\newblock In Camil Demetrescu, editor, {\em Proceedings of the 6th Workshop on
  Experimental Algorithms (WEA'07)}, volume 4525 of {\em Lecture Notes in
  Computer Science}, pages 52--65. Springer, June 2007.

\bibitem{DBLP:conf/gis/DibbeltSW15}
Julian Dibbelt, Ben Strasser, and Dorothea Wagner.
\newblock Fast exact shortest path and distance queries on road networks with
  parametrized costs.
\newblock In Jie Bao, Christian Sengstock, Mohammed~Eunus Ali, Yan Huang,
  Michael Gertz, Matthias Renz, and Jagan Sankaranarayanan, editors, {\em
  Proceedings of the 23rd {SIGSPATIAL} International Conference on Advances in
  Geographic Information Systems, Bellevue, WA, USA, November 3-6, 2015}, pages
  66:1--66:4. {ACM}, 2015.
\newblock \href {https://doi.org/10.1145/2820783.2820856}
  {\path{doi:10.1145/2820783.2820856}}.

\bibitem{dsw-cch-15}
Julian Dibbelt, Ben Strasser, and Dorothea Wagner.
\newblock {Customizable Contraction Hierarchies}.
\newblock {\em ACM Journal of Experimental Algorithmics}, 21(1):1.5:1--1.5:49,
  April 2016.

\bibitem{d-ntpcg-59}
Edsger~W. Dijkstra.
\newblock {A Note on Two Problems in Connexion with Graphs}.
\newblock {\em Numerische Mathematik}, 1(1):269--271, 1959.

\bibitem{d-aassp-69}
Stuart~E. Dreyfus.
\newblock {An Appraisal of Some Shortest-Path Algorithms}.
\newblock {\em Operations Research}, 17(3):395--412, 1969.

\bibitem{DBLP:conf/aaai/EisnerFS11}
Jochen Eisner, Stefan Funke, and Sabine Storandt.
\newblock Optimal route planning for electric vehicles in large networks.
\newblock In Wolfram Burgard and Dan Roth, editors, {\em Proceedings of the
  Twenty-Fifth {AAAI} Conference on Artificial Intelligence, {AAAI} 2011, San
  Francisco, California, USA, August 7-11, 2011}. {AAAI} Press, 2011.

\bibitem{fns-opca-14}
Stefan Funke, Andr{\'e} Nusser, and Sabine Storandt.
\newblock {On k-Path Covers and their Applications}.
\newblock In {\em Proceedings of the 40th International Conference on Very
  Large Databases (VLDB 2014)}, pages 893--902, 2014.

\bibitem{gks-rpfof-10}
Robert Geisberger, Moritz Kobitzsch, and Peter Sanders.
\newblock {Route Planning with Flexible Objective Functions}.
\newblock In {\em Proceedings of the 12th Workshop on Algorithm Engineering and
  Experiments (ALENEX'10)}, pages 124--137. SIAM, 2010.

\bibitem{gssv-erlrn-12}
Robert Geisberger, Peter Sanders, Dominik Schultes, and Christian Vetter.
\newblock {Exact Routing in Large Road Networks Using Contraction Hierarchies}.
\newblock {\em Transportation Science}, 46(3):388--404, August 2012.

\bibitem{gv-errnt-11}
Robert Geisberger and Christian Vetter.
\newblock {Efficient Routing in Road Networks with Turn Costs}.
\newblock In Panos~M. Pardalos and Steffen Rebennack, editors, {\em Proceedings
  of the 10th International Symposium on Experimental Algorithms (SEA'11)},
  volume 6630 of {\em Lecture Notes in Computer Science}, pages 100--111.
  Springer, 2011.

\bibitem{gh-cspas-05}
Andrew~V. Goldberg and Chris Harrelson.
\newblock {Computing the Shortest Path: {A*} Search Meets Graph Theory}.
\newblock In {\em Proceedings of the 16th Annual {ACM--SIAM} Symposium on
  Discrete Algorithms (SODA'05)}, pages 156--165. SIAM, 2005.

\bibitem{gkw-blwr-07}
Andrew~V. Goldberg, Haim Kaplan, and Renato~F. Werneck.
\newblock {Better Landmarks Within Reach}.
\newblock In Camil Demetrescu, editor, {\em Proceedings of the 6th Workshop on
  Experimental Algorithms (WEA'07)}, volume 4525 of {\em Lecture Notes in
  Computer Science}, pages 38--51. Springer, June 2007.

\bibitem{gw-cppsp-05}
Andrew~V. Goldberg and Renato~F. Werneck.
\newblock {Computing Point-to-Point Shortest Paths from External Memory}.
\newblock In {\em Proceedings of the 7th Workshop on Algorithm Engineering and
  Experiments (ALENEX'05)}, pages 26--40. SIAM, 2005.

\bibitem{hs-gbpo-18}
Michael Hamann and Ben Strasser.
\newblock {Graph Bisection with Pareto Optimization}.
\newblock {\em ACM Journal of Experimental Algorithmics}, 23(1):1.2:1--1.2:34,
  2018.
\newblock URL: \url{http://doi.acm.org/10.1145/3173045}.

\bibitem{hnr-afbhd-68}
Peter~E. Hart, Nils Nilsson, and Bertram Raphael.
\newblock {A Formal Basis for the Heuristic Determination of Minimum Cost
  Paths}.
\newblock {\em IEEE Transactions on Systems Science and Cybernetics},
  4:100--107, 1968.

\bibitem{klsv-dtdch-10}
Tim Kieritz, Dennis Luxen, Peter Sanders, and Christian Vetter.
\newblock {Distributed Time-Dependent Contraction Hierarchies}.
\newblock In Paola Festa, editor, {\em Proceedings of the 9th International
  Symposium on Experimental Algorithms (SEA'10)}, volume 6049 of {\em Lecture
  Notes in Computer Science}, pages 83--93. Springer, May 2010.

\bibitem{kswz-erptd-p-20}
Alexander Kleff, Frank Schulz, Jakob Wagenblatt, and Tim Zeitz.
\newblock {Efficient Route Planning with Temporary Driving Bans, Road Closures,
  and Rated Parking Areas.}
\newblock In Simone Faro and Domenico Cantone, editors, {\em Proceedings of the
  18th International Symposium on Experimental Algorithms (SEA'20)}, volume 160
  of {\em Leibniz International Proceedings in Informatics}, 2020.
\newblock URL: \url{https://doi.org/10.4230/LIPIcs.SEA.2020.17}.

\bibitem{DBLP:conf/nips/LikhachevGT03}
Maxim Likhachev, Geoffrey~J. Gordon, and Sebastian Thrun.
\newblock Ara*: Anytime a* with provable bounds on sub-optimality.
\newblock In Sebastian Thrun, Lawrence~K. Saul, and Bernhard Sch{\"{o}}lkopf,
  editors, {\em Advances in Neural Information Processing Systems 16 [Neural
  Information Processing Systems, {NIPS} 2003, December 8-13, 2003, Vancouver
  and Whistler, British Columbia, Canada]}, pages 767--774. {MIT} Press, 2003.

\bibitem{bingblog}
Bing maps new routing engine.
\newblock
  \url{https://blogs.bing.com/maps/2012/01/05/bing-maps-new-routing-engine/}.
\newblock Accessed: 2020-01-25.

\bibitem{ndls-bastd-12}
Giacomo Nannicini, Daniel Delling, Leo Liberti, and Dominik Schultes.
\newblock {Bidirectional {A*} Search on Time-Dependent Road Networks}.
\newblock {\em Networks}, 59:240--251, 2012.
\newblock Best Paper Award.

\bibitem{or-tnp-89}
Ariel Orda and Raphael Rom.
\newblock {Traveling without waiting in time-dependent networks is NP-hard}.
\newblock Technical report, Dept. Electrical Engineering, Technion-Israel
  Institute of Technology, 1989.

\bibitem{swz-umlgt-02}
Frank Schulz, Dorothea Wagner, and Christos Zaroliagis.
\newblock {Using Multi-Level Graphs for Timetable Information in Railway
  Systems}.
\newblock In {\em Proceedings of the 4th Workshop on Algorithm Engineering and
  Experiments (ALENEX'02)}, volume 2409 of {\em Lecture Notes in Computer
  Science}, pages 43--59. Springer, 2002.

\bibitem{DBLP:conf/socs/StrasserHB14}
Ben Strasser, Daniel Harabor, and Adi Botea.
\newblock Fast first-move queries through run-length encoding.
\newblock In Stefan Edelkamp and Roman Bart{\'{a}}k, editors, {\em Proceedings
  of the Seventh Annual Symposium on Combinatorial Search, {SOCS} 2014, Prague,
  Czech Republic, 15-17 August 2014}. {AAAI} Press, 2014.

\bibitem{swz-sfert-20}
Ben Strasser, Dorothea Wagner, and Tim Zeitz.
\newblock {Space-efficient, Fast and Exact Routing in Time-dependent Road
  Networks}.
\newblock In {\em Proceedings of the 28th Annual European Symposium on
  Algorithms (ESA'20)}, Leibniz International Proceedings in Informatics,
  September 2020.

\bibitem{strasser2019perfect}
Ben Strasser and Tim Zeitz.
\newblock A* with perfect potentials, 2019.
\newblock \href {http://arxiv.org/abs/1910.12526} {\path{arXiv:1910.12526}}.

\bibitem{DBLP:conf/socs/SturtevantTTUKS15}
Nathan~R. Sturtevant, Jason~M. Traish, James~R. Tulip, Tansel Uras, Sven
  Koenig, Ben Strasser, Adi Botea, Daniel Harabor, and Steve Rabin.
\newblock The grid-based path planning competition: 2014 entries and results.
\newblock In Levi Lelis and Roni Stern, editors, {\em Proceedings of the Eighth
  Annual Symposium on Combinatorial Search, {SOCS} 2015, 11-13 June 2015, Ein
  Gedi, the Dead Sea, Israel}, page 241. {AAAI} Press, 2015.

\bibitem{t-dfslg2-72}
Robert Tarjan.
\newblock {Depth-First Search and Linear Graph Algorithms}.
\newblock {\em SIAM Journal on Computing}, 1972.

\bibitem{DBLP:conf/aaai/UrasK14}
Tansel Uras and Sven Koenig.
\newblock Identifying hierarchies for fast optimal search.
\newblock In Carla~E. Brodley and Peter Stone, editors, {\em Proceedings of the
  Twenty-Eighth {AAAI} Conference on Artificial Intelligence, July 27 -31,
  2014, Qu{\'{e}}bec City, Qu{\'{e}}bec, Canada}, pages 878--884. {AAAI} Press,
  2014.

\bibitem{DBLP:conf/aaai/ZhouH02}
Rong Zhou and Eric~A. Hansen.
\newblock Multiple sequence alignment using anytime a\({}^{\mbox{*}}\).
\newblock In Rina Dechter, Michael~J. Kearns, and Richard~S. Sutton, editors,
  {\em Proceedings of the Eighteenth National Conference on Artificial
  Intelligence and Fourteenth Conference on Innovative Applications of
  Artificial Intelligence, July 28 - August 1, 2002, Edmonton, Alberta,
  Canada}, pages 975--977. {AAAI} Press / The {MIT} Press, 2002.

\end{thebibliography}
\end{document}